\documentclass[submission,copyright,creativecommons]{eptcs}
\providecommand{\event}{Workshop on Models for Formal Analysis of Real Systems (MARS 2022)} % Name of the event you are submitting to

\usepackage{comment}
\usepackage{enumitem}
\usepackage{subfig}
\usepackage{graphicx}
\usepackage[export]{adjustbox}
\usepackage{multirow}
\usepackage[table]{xcolor}
\usepackage{tikz}
\usepackage{bm}
\usepackage{amssymb}
\usepackage{amsmath}

\usepackage{custom-packages/uppaal-latex-master/uppaal}

\usepackage{hyperref}
\usepackage{cleveref}

\crefname{section}{Sec.}{Sec.}
\Crefname{section}{Section}{Sections}
\crefname{subsection}{Sec.}{Sec.}
\Crefname{subsection}{Section}{Sections}
\crefname{subsubsection}{Sec.}{Sec.}
\Crefname{subsubsection}{Section}{Sections}

\crefname{theorem}{Thm.}{Thm.}
\Crefname{theorem}{Theorem}{Theorems}
\crefname{lemma}{Lem.}{Lem.}
\Crefname{lemma}{Lemma}{Lemmata}
\crefname{corollary}{Cor.}{Cor.}
\Crefname{corollary}{Corollary}{Corollaries}
\crefname{proposition}{Prop.}{Prop.}
\Crefname{proposition}{Proposition}{Propositions}
\crefname{fact}{Fact}{Facts}
\Crefname{fact}{Fact}{Facts}
\crefname{definition}{Def.}{Def.}
\Crefname{definition}{Definition}{Definitions}

\crefname{example}{Ex.}{Ex.}
\Crefname{example}{Example}{Examples}
\crefname{remark}{Rm.}{Rm.}
\Crefname{remark}{Remark}{Remarks}
\crefname{equation}{Eq.}{Eq.}
\Crefname{equation}{Equation}{Equations}

\crefname{figure}{Fig.}{Fig.}
\Crefname{figure}{Figure}{Figures}
\crefname{table}{Tab.}{Tab.}
\Crefname{table}{Table}{Tables}
\crefname{listing}{\lstlistingname}{\lstlistingname}
\Crefname{listing}{Listing}{Listings}

\title{Modeling $\mathbb{R}^3$ Needle Steering in Uppaal}

\author{
Sascha Lehmann$^{1}$ \qquad Antje Rogalla$^{1}$ \qquad Maximilian Neidhardt$^{2}$ \\
Anton Reinecke$^{1}$ \qquad Alexander Schlaefer$^{2}$ \qquad Sibylle Schupp$^{1}$
\institute{
$^{1}$Institute for Software Systems
\qquad
$^{2}$Institute of Medical Technology and Intelligent Systems\\Hamburg University of Technology, Hamburg, Germany\thanks{This study was partially funded by the TUHH i$^3$~lab initiative (T-LP-E01-WTM-1801-02), DFG SCHU 2479, and DFG SCHL 1844/6-1.}}
\email{\{s.lehmann, antje.rogalla, maximilian.neidhardt, anton.reinecke, schlaefer, schupp\}@tuhh.de}
}

\def\titlerunning{Modeling 3D Needle Steering}
\def\authorrunning{S. Lehmann et al.}

\begin{document}
\maketitle

%%%%%%%%%%%%%%%%%%%%%%%%%%%%%%%%%%%%%%%%%%%%%%%%%%%%%%%%%%%%%%%%%%%%%%%%%%%%%%%%
%%% Abstract %%%
%%%%%%%%%%%%%%%%%%%%%%%%%%%%%%%%%%%%%%%%%%%%%%%%%%%%%%%%%%%%%%%%%%%%%%%%%%%%%%%%
\begin{abstract}
  Medical cyber-physical systems are safety-critical, and as such, require ongoing verification of their correct behavior, as system failure during run time may cause severe (or even fatal) personal damage.
  However, creating a verifiable model often conflicts with other application requirements, most notably regarding data precision and model accuracy, as efficient model checking promotes discrete data (over continuous) and abstract models to reduce the state space.
  In this paper, we approach the task of medical needle steering in soft tissue around potential obstacles.
  We design a verifiable model of needle motion (implemented in Uppaal Stratego) and a framework embedding the model for online needle steering.
  We mitigate the conflict by imposing boundedness on both the data types, reducing from $\mathbb{R}^3$ to $\mathbb{Z}^3$ when needed, and the motion and environment models, reducing the set of allowed local actions and global paths.
  In experiments, we successfully apply the static model alone, as well as the dynamic framework in scenarios with varying environment complexity and both a virtual and real needle setting, where up to $100\%$ of targets were reached depending on the scenario and needle.
\end{abstract}

%%%%%%%%%%%%%%%%%%%%%%%%%%%%%%%%%%%%%%%%%%%%%%%%%%%%%%%%%%%%%%%%%%%%%%%%%%%%%%%%
%%% Introduction %%%
%%%%%%%%%%%%%%%%%%%%%%%%%%%%%%%%%%%%%%%%%%%%%%%%%%%%%%%%%%%%%%%%%%%%%%%%%%%%%%%%
\section{Introduction} \label{sec:introduction}
Medical needle steering describes the task of steering flexible and beveled needles in soft tissue towards a target e.g., for a biopsy or brachytherapy, while avoiding critical tissue layers and organs.
The underlying problem is thus the safe navigation of a controllable entity through a (partially known and uncontrollable) environment in $\mathbb{R}^3$.
The following requirements arise for a model intended to approach this underlying problem:
\begin{itemize}[noitemsep, leftmargin=*]
  \item[] \textbf{Safety:} To avoid severe or even fatal damage to a patient by piercing critical tissue or organs, the system needs to behave safely at any time.
  \item[] \textbf{Verifiability:} To ensure safety by proving that no critical state is ever reached, the model needs to be formally verifiable.
  \item[] \textbf{Model Accuracy:} To relate the model to the underlying system, a suitable modeling of the $\mathbb{R}^3$ space, needle motion, and environments is required which reflect all characteristics that could threaten safety.
  \item[] \textbf{Precision:} To limit the deviation of the model and real system, sufficient precision is required in terms of data measurement and data storing in digital systems, which enforce a discretization of data that is naturally continuous in reality.
  \item[] \textbf{Performance:} To use a model in medical cyber-physical systems, real-time performance of adapting and checking the model is required locally for each system step (as system and model would drift apart otherwise), and furthermore, the desired global result should be reached in suitable time (as medical interventions cannot take arbitrary amounts of time).
\end{itemize}
While some requirements are compatible and interdependent (e.g., safety and verifiability), others clearly contradict each other.
In particular, model accuracy and precision for $\mathbb{R}^3$ spaces conflict with performance requirements, as realistic and physically accurate modeling with high precision data usually leads to strongly growing state spaces and calculation times, which in turn need to be bounded to meet constraints of the real system (e.g., feasible durations of surgeries, anesthesia, breath holding, etc. for medical applications).
Furthermore, verifiability conflicts with model accuracy and precision, as model checkers naturally require abstractions from the continuous $\mathbb{R}^3$ domain of the underlying real system (or its fine-grained discretization in $\mathbb{Z}^3$ imposed by limited sensor resolution) to discrete data types supported by the checker.
The key aspect for such models is thus to find a suitable balance between these requirements to make the model both verifiable in reasonable time and relatable to the real system.
Only then, one can guarantee safety both locally (via model checking of a static system snapshot) and globally (via system adaptions whenever safety might become threatened) for a dynamic system like needle steering.

In this paper, we design a model and framework that can be used for the task of online needle steering.
Based on matrix and vector calculations of circle motions, the model implements a geometric model in $\mathbb{R}^3$ space, which was shown to be sufficiently approximating for needle motion in literature \cite{Webster2006}.
In particular, we make the following contributions:
\begin{enumerate}
  \item We provide a base model of needle steering as timed automaton in Uppaal, which adheres to the aforementioned requirements of safety, verifiability, accuracy, precision, and performance to an extent that allows online $\mathbb{R}^3$ needle steering.
  \item We cover design decisions for model simplifications regarding the data domain, entities, and actions, which impose different degrees of boundedness to the data types and motion model.
  \item We provide a framework which the model is embedded into for online strategy synthesis (OSS) on top of Uppaal.
  \item We perform experiments on multiple environment settings with both a virtual and a real needle to show applicability and requirement compliance in practice.
\end{enumerate}

The paper and model build on and extend our previous work on needle steering and OSS.
In ``Synthesizing Strategies for Needle Steering in Gelatin Phantoms'' \cite{Rogalla2020}, we provided a basic model for matching of $\mathbb{R}^2$-projected needle traces against a $\mathbb{R}^2$ motion model via offline strategy synthesis.
In the current paper, we extend the model to $\mathbb{R}^3$ motion to fully reflect the real-world use case of needle steering, which all the more raises the question for sufficiently precise domain modeling to support formal verification in reasonable time.
Furthermore, we elevate the use case from \textit{offline} matching to \textit{online} navigation by embedding the model into the OSS framework to react to environmental uncertainties and the fact that tissue in reality is inhomogeneous and has anatomic obstacles.
``Online Strategy Synthesis for Safe and Optimized Control of Steerable Needles'' \cite{Lehmann2021} introduces the concept of OSS, but does not cover the specifics of the underlying model and modeling decisions, which are subject to the current paper.

The paper is structured as follows:
We present preliminaries on strategy synthesis in \cref{sec:preliminaries}.
Then, we introduce the Uppaal model of needle steering in \cref{sec:needle-steering-model} and explain the OSS framework and the embedding of the Uppaal model in \cref{sec:oss-framework}.
Afterwards, we perform experiments with both a virtual and a real needle in \cref{sec:experiments}.
Finally, we describe the related work in \cref{sec:related-work} and conclude our work in \cref{sec:conclusion}.

%%%%%%%%%%%%%%%%%%%%%%%%%%%%%%%%%%%%%%%%%%%%%%%%%%%%%%%%%%%%%%%%%%%%%%%%%%%%%%%%
%%% Preliminaries %%%
%%%%%%%%%%%%%%%%%%%%%%%%%%%%%%%%%%%%%%%%%%%%%%%%%%%%%%%%%%%%%%%%%%%%%%%%%%%%%%%%
\section{Preliminaries} \label{sec:preliminaries}

\begin{figure}[t]
  \subfloat[The experiment setup]{
  \begin{minipage}[b]{0.45\textwidth}
     \centering
     \includegraphics[width=1.0\linewidth]{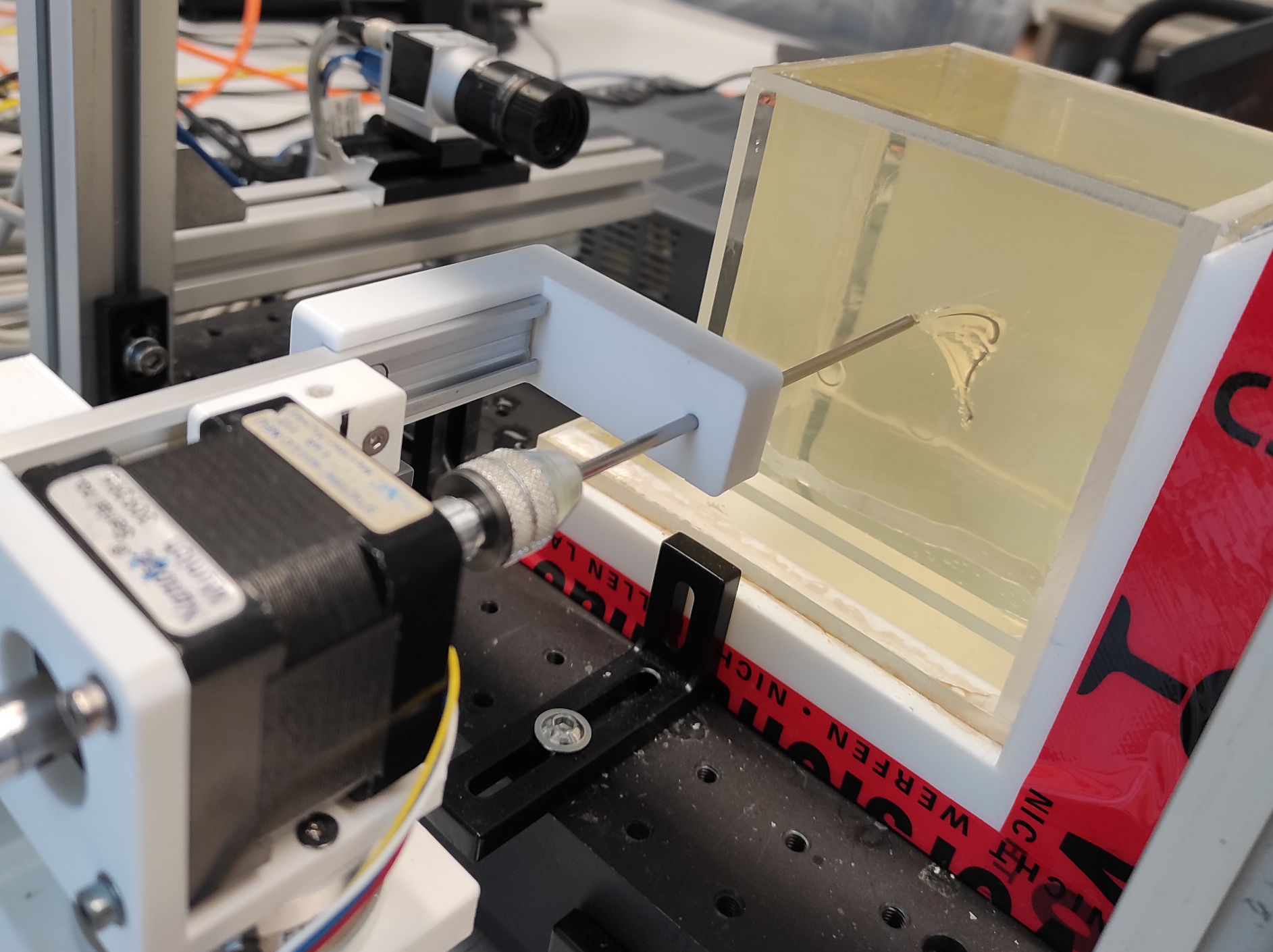}
     \label{subfig:full-setup-view}
  \end{minipage}}
  \hfill
  \subfloat[The bent needle inside a gelatin phantom]{
  \begin{minipage}[b]{0.45\textwidth}
     \centering
     \includegraphics[width=1.0\linewidth]{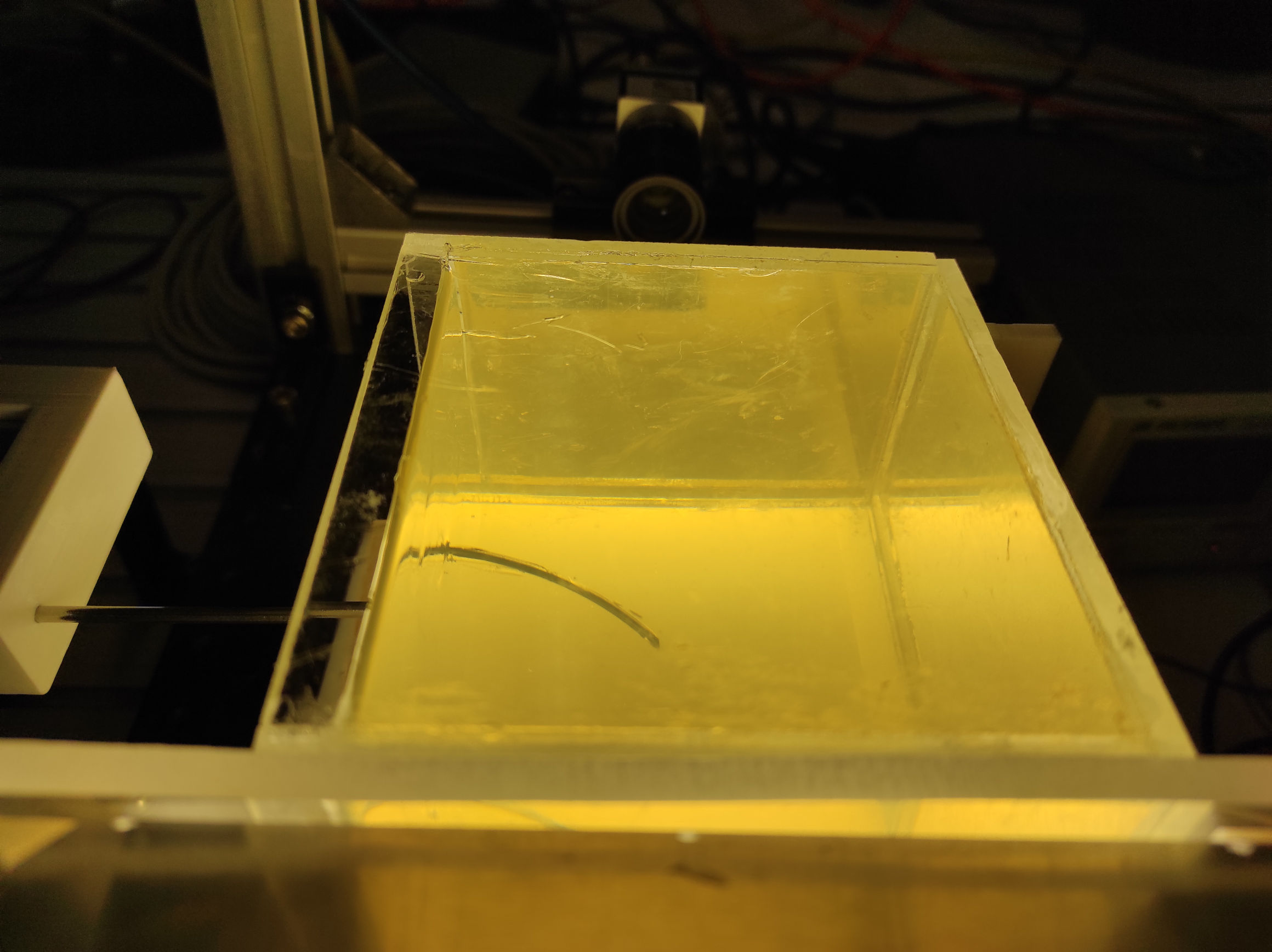}
     \label{subfig:needle-view}
  \end{minipage}}
  % \hfill
  \caption{The experiment setup for needle steering. (a) shows the setup with stepper motor (left), needle guide (center), and gelatin phantom (right). (b) shows the trajectory of a needle moved through a gelatin phantom without rotations.}
  \label{fig:system-setup}
\end{figure}

%%%%%%%%%%%%%%%%%%%%%%%%%%%%

\paragraph{Needle Steering}
Placing long and thin biopsy needles in deep tissue structures is a common medical intervention.
To increase the accuracy in reaching a desired target the physician directs the needle with image guidance, retracting and re-orientating the needle axis multiple times during insertions if needed.
This leads to increased trauma and interventions longer than necessary.
The objective of automated needle steering thus is twofold: to move the needle as closely and directly as possible to the deformable target, and to bypass nerves or any other anatomic obstacles along that path.
The system setup using gelatin phantoms as tissue, as it is common in robotic tissue simulations, is shown in \cref{fig:system-setup}.

Moving needles with robotic guidance through tissue in a controlled way is no trivial task.
The position of the needle depends not only on its velocity and force, and the exact shape of its tip, but also on parameters that are unknown, e.g., the (in-)homogeneity of the surrounding tissue and its elasticity.
Inhomogeneity can, for example, result in abrupt needle movements and subsequent overshoots;
elasticity determines the forces needed.
Moreover, in brachytherapy or anesthesia clinical targets are often not located directly under the skin and cannot be reached with stiff needles but require flexible needles with bevel tips.
Yet flexible needles complicate the computation of needle movements further since the force between tissue and needle tip can now result in small side-wards movements.
Altogether, the needle-tissue interaction is hard to define and predict, and presents to date the major challenge in automating the process of needle steering.

\paragraph{Environment Setting}
The environment is a $\mathbb{R}^3$ space, and interpreted as a set of regions:
Target regions (TR) represent the sections where we want to navigate to, critical regions (CR) model unsafe regions which we want to circumvent (e.g., organs), and detection regions (DR) model the surrounding of CRs in which we can detect a nearing CR (e.g., by measuring increasing force).
Furthermore, unknown regions (UR) model sections which were not discovered yet, and safe regions (SR) represent all sections that are guaranteed safe, including both TRs and DRs.
See \cite{Lehmann2021} for further details on the region interpretation.

\paragraph{Uppaal Stratego}
The \textit{Uppaal Stratego} tool \cite{David2015b} is an integrated tool suite for generation and optimization of strategies.
It combines the statistical functionalities of Uppaal SMC \cite{David2015} with the strategy synthesis capabilities of Uppaal Tiga \cite{Behrmann2007}, and uses stochastic priced timed games as base model formalism.
In particular, the tool guarantees to generate a provably correct and complete winning strategy (whenever one exists) for a controllable entity in a two-player game against an uncontrollable environment.
We use the tool for the offline strategy synthesis step inside the OSS framework.

%%%%%%%%%%%%%%%%%%%%%%%%%%%%%%%%%%%%%%%%%%%%%%%%%%%%%%%%%%%%%%%%%%%%%%%%%%%%%%%%
%%% The Needle Steering Model %%%
%%%%%%%%%%%%%%%%%%%%%%%%%%%%%%%%%%%%%%%%%%%%%%%%%%%%%%%%%%%%%%%%%%%%%%%%%%%%%%%%
\section{The Needle Steering Model} \label{sec:needle-steering-model}

\subsection{Components} \label{subsec:model-components}

\begin{figure}[t]
  \begin{minipage}[b]{0.37\textwidth}
    \centering
    \subfloat[Decision Maker / User]{
       \centering
       \includegraphics[width=0.95\linewidth]{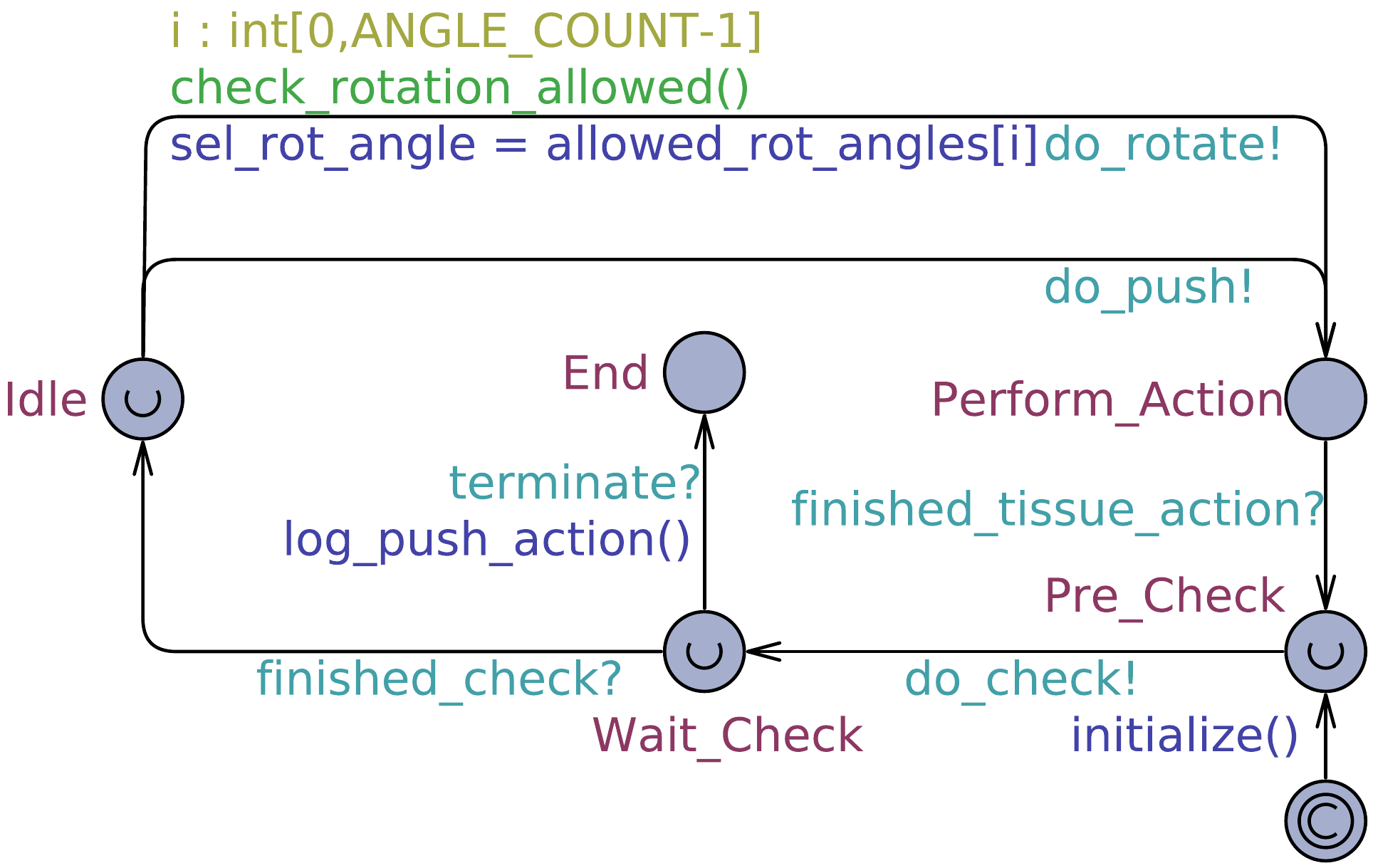}
       \label{subfig:user-template}
    }
    \hfill
    \subfloat[Controlled Device / Needle]{
       \centering
       \includegraphics[width=0.95\linewidth]{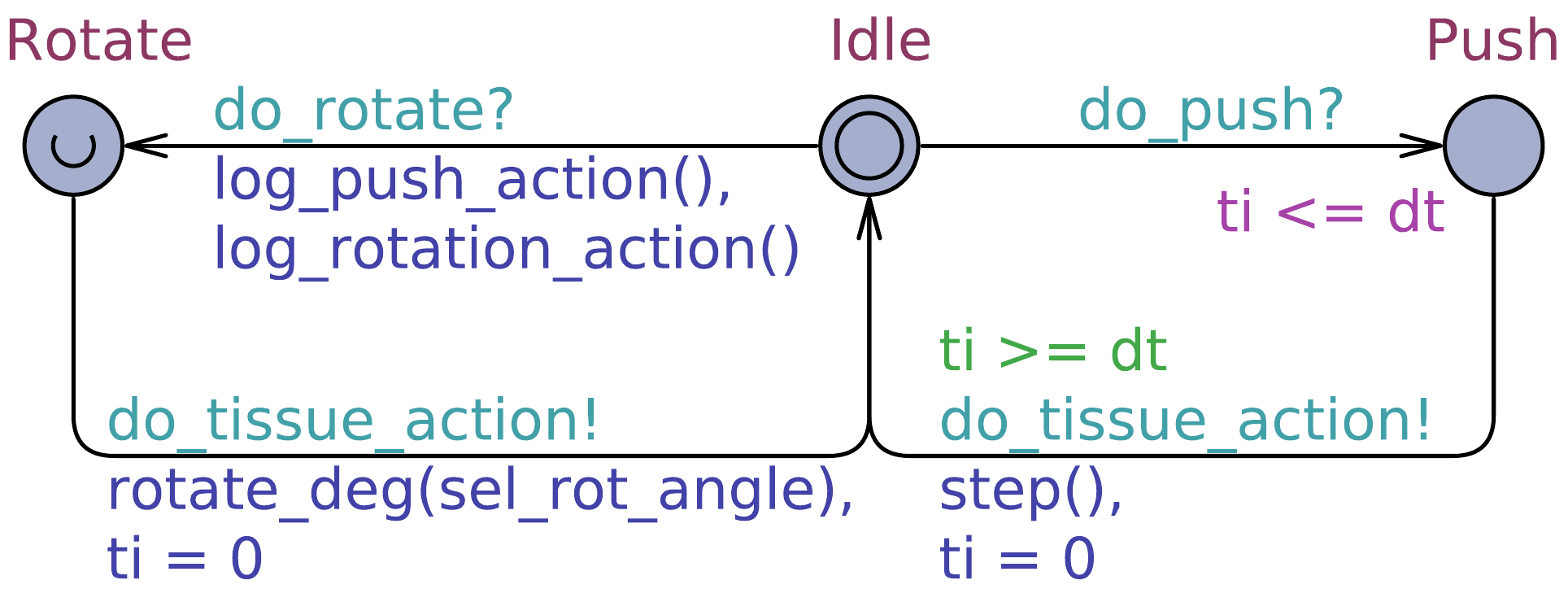}
       \label{subfig:needle-template}
    }
    \hfill
    \subfloat[Environment / Tissue]{
       \centering
       \includegraphics[width=0.65\linewidth]{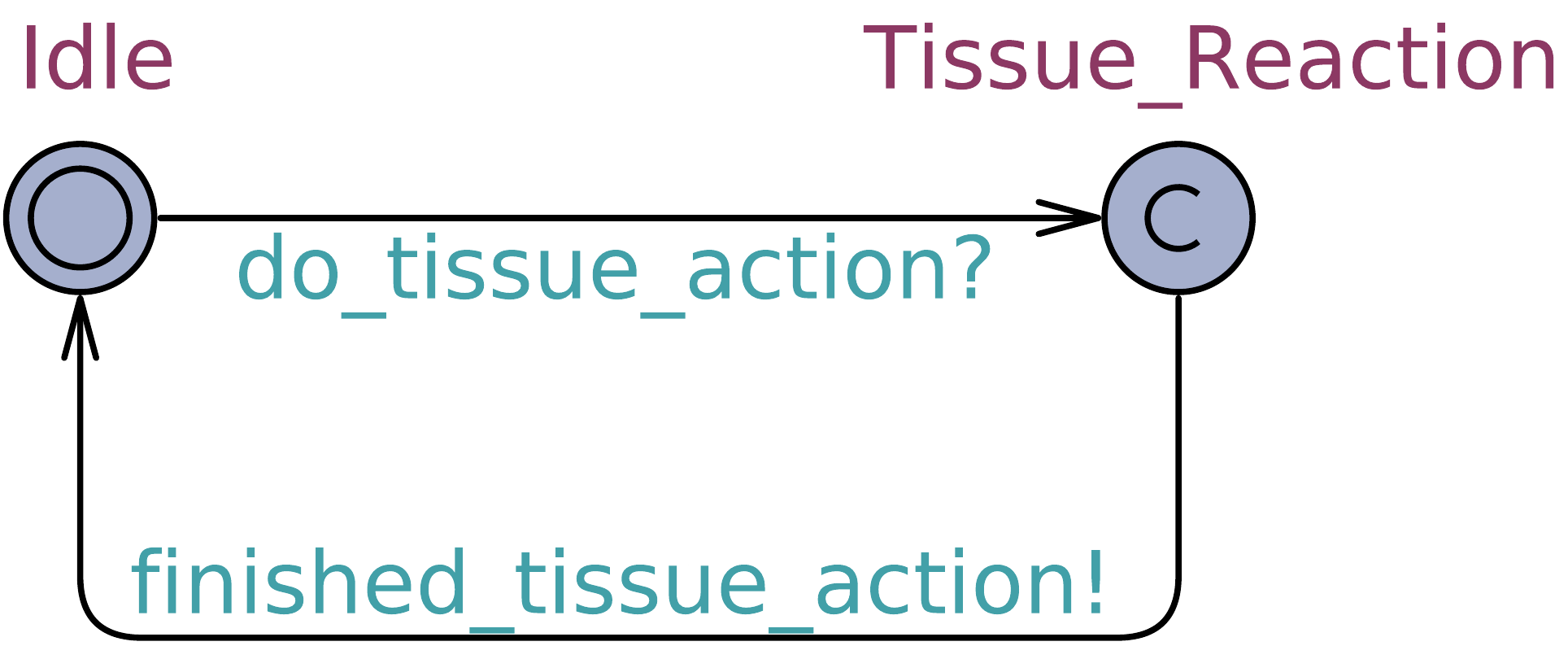}
       \label{subfig:tissue-template}
    }
  \end{minipage}
  \hfill
  \begin{minipage}[b]{0.58\textwidth}
    \subfloat[State Checker]{
       \centering
       \includegraphics[width=0.95\linewidth]{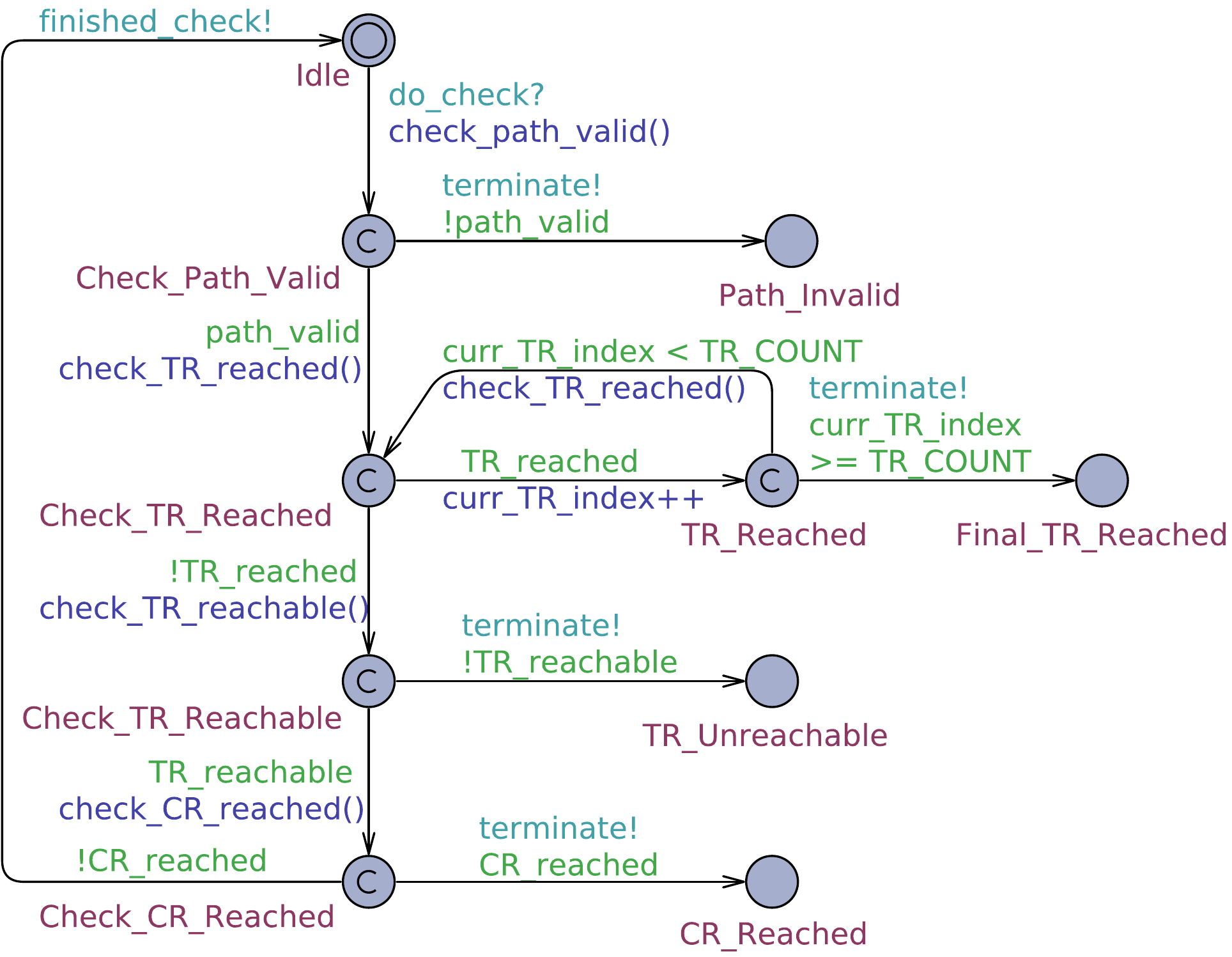}
       \label{subfig:state-checker-template}
    }
  \end{minipage}
  \hspace{0.1cm}
  \caption{The Uppaal model components}
  \label{fig:uppaal-model-templates}
\end{figure}

The three main component types in the Uppaal model are concrete \textit{entity models}, abstract \textit{process models}, and the underlying \textit{C code declaration}.
The model components are shown in \cref{fig:uppaal-model-templates}.

Three entity models exist in total:
The \texttt{Decision Maker} (\cref{subfig:user-template}) represents the clinical expert, who initiates actions of the device.
The \texttt{Controlled Device} (\cref{subfig:needle-template}) represents the needle, which receives instructions from the user and performs the actions accordingly.
The \texttt{Environment} (\cref{subfig:tissue-template}) represents the tissue, which may react to changing needle states, or adapt due to external influences over time.

Furthermore, one process model exists:
The \texttt{State Checker} (\cref{subfig:state-checker-template}) performs checks on each model state to ensure validity of the current trace, and ``discards'' the trace (by leading to a deadlock state) as soon as validity is violated or the target is reached.

\subsection{Execution Flow} \label{subsec:model-execution-flow}
The overall execution flow of the model is as follows:
The \texttt{User} initializes the system (\uppPlain{initialize()}) and initiates a check of the initial system state (\uppPlain{do_check!}).
The \texttt{State Checker} then traverses the locations \uppPlain{Check_Path_Valid}, \uppPlain{Check_TR_Reached}, \uppPlain{Check_TR_Reachable}, and \uppPlain{Check_CR_Reached} where the four functions \uppPlain{check_path_valid()}, \uppPlain{check_TR_reached()}, \uppPlain{check_TR_reachable()}, and \uppPlain{check_CR_reached()} check whether the current needle path is valid, the current TR is reached or at least still reachable, or any CR is currently reached, respectively.
In case that a TR is reached, the system switches to the next TR (\uppPlain{curr_TR_index++}).
If either a path is determined as invalid, the final TR is reached, the current TR is not reachable anymore, or a CR is reached, the path is successfully classified and needs no further evaluation, so that the \texttt{State Checker} switches to \uppPlain{Path_Invalid}, \uppPlain{Final_TR_Reached}, \uppPlain{TR_Unreachable}, or \uppPlain{UR_Reached}, respectively, and the \texttt{User} switches to \uppPlain{End};
these states are intended as deadlocks.

If the \texttt{State Checker} is successfully traversed, the \texttt{User} provides an instruction for the \texttt{Needle} on the \uppPlain{Idle -> Perform_Action} edge, which is either a rotation (\uppPlain{do_rotate!}) or a push motion (\uppPlain{do_push!}).
The instruction then enables the \texttt{Needle}, which performs the corresponding action via \uppPlain{rotate_deg()} or \uppPlain{step()}, which might include particular time delays (e.g., \uppPlain{ti <= dt} and \uppPlain{ti >= dt} in \uppPlain{Push}) or auxiliary logging routines (e.g., \uppPlain{log_push_action()} and \uppPlain{log_rotation_action()} on \uppPlain{Idle -> Rotate}).

Once the action step is successfully executed, the \uppPlain{do_tissue_action!} call enables the \texttt{Tissue} model, which may then react to the action of the controllable entity.
In \cref{subfig:tissue-template}, the simplest form of tissue model is shown, which performs no further reaction, i.e., keeps the CRs and TRs static;
uncontrollable actions, such as stochastical or periodical movement CRs and TRs, can then be added to this model depending on the specifics of the tissue.
Finally, the \texttt{Tissue} model calls \uppPlain{finished_tissue_action!} to signalize to the \texttt{User} that the reaction has finished, and the \texttt{User} starts the next cycle of state checking, instruction, action.

\subsection{C Code Declaration and Queries} \label{subsec:model-c-code-declaration}
The introduced network of automata implements the integration of components, the synchronization between individual submodels, and the overall flow and timing of actions and reactions.
The underlying code used by these models is defined in the Uppaal C code declaration section.
The complete C code declaration can be found in \cref{app:c-code-declaration}, and provides the following:
\begin{itemize}
  \item The \textit{data representation of all entities}, which includes the position, rotation and velocity of the needle, as well as the positions and sizes of regions (i.e., DRs, CRs, TRs).
  \item The \textit{mathematical structures} such as matrices and vectors in $\mathbb{Z}^3$ required for geometric calculations.
  \item The \textit{algebraic operations} in $\mathbb{R}^3$ on the aforementioned structures, such as products of matrices and vectors, vector normalizations, and magnitude calculations.
  \item The \textit{motion model functions}, such as motion circle calculation, step, and rotate.
  \item The \textit{state checking functions} for TR reaching and reachability, CR reaching, and path validity.
\end{itemize}
Given the complete model, we define the following two queries:
\begin{gather}
  \texttt{E<> Checker.Final\_TR\_Reached} \label{eq:query-exists-path}\\
  \texttt{strategy ReachFinalTR = control: A<> Checker.Final\_TR\_Reached} \label{eq:query-strategy}
\end{gather}

Query \ref{eq:query-exists-path} allows checking whether a path exists that leads to the final TR, and generates a corresponding model trace for review in the Uppaal simulator if configured accordingly.
Query \ref{eq:query-strategy} synthesizes a strategy over all possible paths towards the final TR;
from this strategy, we can extract the set of suitable instruction plans for needle motion.
The later query is frequently called by the framework defined in \cref{sec:oss-framework} during the offline strategy synthesis step.

\subsection{Simplifications} \label{subsec:model-simplifications}
To use the model for motion plan generation and safety guarantees in our application, it needs to conform with the requirements and conflicts introduced in \cref{sec:introduction}.
In fact, modeling in $\mathbb{R}^3$ and allowing any possible motion in that space easily leads to state explosion problems, which would inhibit verifiability even if a verifier supported $\mathbb{R}^3$ data.
However, simple modeling in turn often conflicts with precision and accuracy criteria.
To reduce the state space and balance the requirements to an extent suitable for the needle steering application, we implement $3$ types of simplifications, which impose certain degress of boundedness:
\textit{value} restrictions, \textit{motion} restrictions, and \textit{environment} restrictions.

As value restrictions, we implement a partial discretization of data, i.e., all data is scaled by a factor \uppPlain{_S_} and cast to \texttt{int} for storage, and only cast to \texttt{double} for intermediate calculations.
That way, all states relevant for model checking consist only of bounded data.
Furthermore, we discretize time with a fixed step size \uppPlain{dt} per action step, which can be increased to improve performance at the expense of precision.

As motion restrictions, we limit the type and extent of needle motion.
The \textit{rotate} motion only allows a rotation by $90^{\circ}$, $180^{\circ}$, or $270^{\circ}$, and we set a maximum number of allowed rotations as well as a lower bound of push motion distance required between two consecutive rotations.
Apart from these restrictions, the otherwise arbitrary choice of rotation points still allows navigating to most (if not all) physically feasible regions, and we will see in \cref{sec:experiments} that the target regions are usually reachable under these abstractions.
Furthermore, a \textit{pull} motion is not explicitly implemented in the model, as such motion only becomes relevant in the OSS workflow (see \cref{sec:oss-framework}), and would lead to an infinite state space otherwise, as the model could loop (\textit{push}, \textit{pull}) infinitely often.
The \uppPlain{check_path_valid()} and \uppPlain{check_TR_reachable()} functions limit the state space further by omitting paths that lead to physically impossible needle trajectories or will not be able to reach a TR in the future, respectively.

Finally, as environment restrictions, we limit the complexity of entities and reactions.
In the current model iteration, all regions (DRs, CRs, TRs) are modelled as simple spheres, and the regions remain static over time, i.e., will not move due to unpredictable patient movements or breathing.

%%%%%%%%%%%%%%%%%%%%%%%%%%%%%%%%%%%%%%%%%%%%%%%%%%%%%%%%%%%%%%%%%%%%%%%%%%%%%%%%
%%% The Online Strategy Synthesis Framework %%%
%%%%%%%%%%%%%%%%%%%%%%%%%%%%%%%%%%%%%%%%%%%%%%%%%%%%%%%%%%%%%%%%%%%%%%%%%%%%%%%%
\section{The Online Strategy Synthesis Framework} \label{sec:oss-framework}
The model designed in \cref{sec:needle-steering-model} allows offline verification and strategy synthesis in a static manner, i.e., based on a particular snapshot of the real system.
The environment of a real system is usually not fully known though.
For example, the concrete characteristics of the needle motion circle is initially unknown if needle and tissue cannot be fully characterized a-priori, and we typically do not know whether critical regions exist or where they are situated.
Furthermore, measurements of the real system (e.g., position and force data) may reveal that the needle deviates from the prediction or is about to become unsafe due to increasing forces, or that initially calculated motion circles do not fit observations at a latter stage.
The Uppaal model alone would then not suffice to react to environment changes or dynamically obtained system knowledge.
Fortunately, the model can be leveraged from offline to online by embedding it into a framework that dynamically updates the model with new knowledge, and triggers system adaptions and strategy resynthesis as needed.

In the following, we distinguish between $3$ types of knowledge:
actual, a-priori, and dynamically learned knowledge.
The \textit{actual knowledge} comprises all knowledge obtainable from the system, and is guaranteed to be correct.
The \textit{a-priori knowledge} is usually a subset of actual knowledge, but may also include knowledge that is assumed to be correct beforehand, but turns out to be wrong during the experiment.
For needle steering, such a-priori knowledge can be partial knowledge about CRs, and the default value of the motion circle radius.
The \textit{dynamically learned knowledge} finally contains all knowledge that was not known in advance, but is discovered and learned on-the-fly during the experiment.
Such knowledge is always an assumption, as it is based on the interpretation of observed data.
Examples are the detected DRs and CRs, the current position and force data of the needle, and the needle motion circle derived from the observed position data.

The model alone would only lead to correct results if we had access to the full actual knowledge, and if the system continues to behave exactly like the model.
Such idealized conditions are rarely given, as the tissue is usually inhomogenous and patient-specific, and the needle motion is only approximately circular.
Thus, the framework adds the following functionality on top of the Uppaal model:
\begin{itemize}
  \item real-time data acquisition for model updates
  \item pull-back motion to retreat from discovered CRs (cf. \cref{subsec:model-simplifications})
  \item determination of the actual motion circle (both initially and after each system adaption)
\end{itemize}

\subsection{Workflow} \label{sec:framework-workflow}

\begin{figure}[t]
  \centering
  \includegraphics[width=1.0\linewidth]{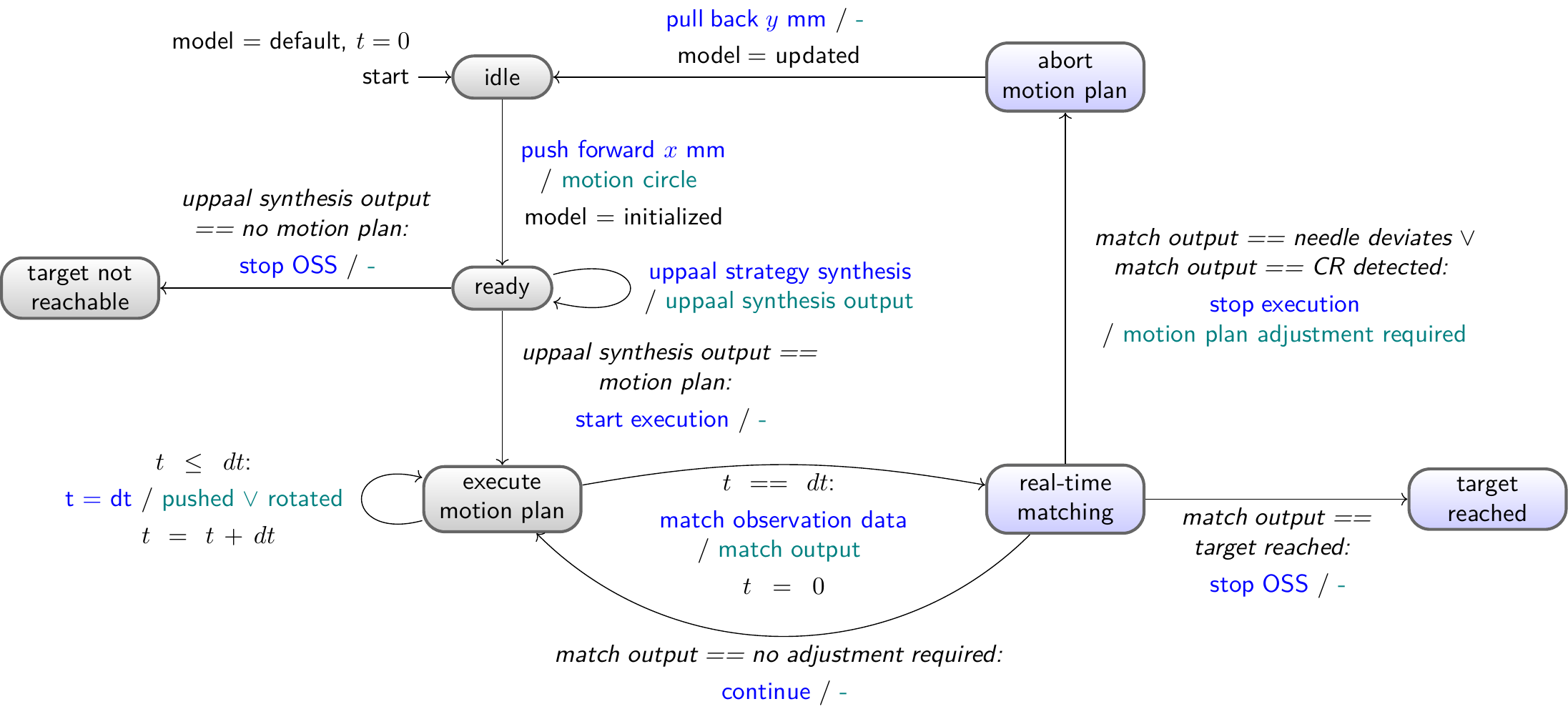}
  \caption{The OSS workflow (grey states covered by offline synthesis, blue states added by framework)}
  \label{fig:oss-workflow}
\end{figure}

The workflow of the framework is shown in \cref{fig:oss-workflow} and covered in \cite{Lehmann2021}, so that we will only summarize the important steps:
First, the needle is pushed slightly, and the measured position data is used to obtain an initial motion circle.
The model is updated accordingly, and an initial strategy to the target is calculated by Uppaal Stratego, from which the framework derives a concrete motion plan.
This motion plan is executed parallel to ongoing measurements of the real system, until the target is reached, or the acquired knowledge contradicts the current assumptions (e.g., the needle deviates, or force exceeds given bounds).
In the later case, the plan is aborted, and the needle is readjusted, followed by the recalculation of the motion circle, strategy, and motion plan, and execution of the latter.
If no strategy can be obtained even after all possible readjustments (i.e., when the needle has been pulled back to the start), the user is informed accordingly.

\subsection{Simplifications} \label{sec:framework-simplifications}
Simplifications, similar to the ones on model level (cf. \cref{subsec:model-simplifications}), are required on framework level to meet the application requirements and deal with their conflicts.
While the model simplifications were mostly concerned with verifiability, accuracy, and precision requirements, the framework simplification mostly targets performance aspects and the weakening of safety checks whenever allowed during the process.
We again distinguish between value, motion, and environment restrictions.
As value restrictions, we only consider measured data points at a particular time resolution, i.e., use one data point every $n$ time units and discard the rest.
As motion restrictions, we always readjust the needle by a fixed pull and push distance, and do not calculate dynamic readjustment distances based on the assumed sizes of detected CRs.
Finally, as environment restrictions, we perform no checks for deviation or CRs during pull-back motions.
We argue that neither deviation can be exceeded nor CRs can be reached at that moment, as the pull-back motion only follows the formerly pierced path, and the presence of the needle prevents CRs to move into these positions, respectively.

Recalling the limited allowed number of $n$ rotations, which we introduced as a simplification inside the model to reduce the state space, we note that this limit is applied only on model level, so that at each strategy resynthesis step, another $n$ rotations are allowed.

%%%%%%%%%%%%%%%%%%%%%%%%%%%%%%%%%%%%%%%%%%%%%%%%%%%%%%%%%%%%%%%%%%%%%%%%%%%%%%%%
%%% Experiments %%%
%%%%%%%%%%%%%%%%%%%%%%%%%%%%%%%%%%%%%%%%%%%%%%%%%%%%%%%%%%%%%%%%%%%%%%%%%%%%%%%%
\section{Experiments} \label{sec:experiments}

\begin{figure}[t]
  \centering
  \hfill
  \subfloat[No\_CR]{
  \begin{minipage}[b]{0.30\textwidth}
     \centering
     \includegraphics[width=1.0\linewidth]{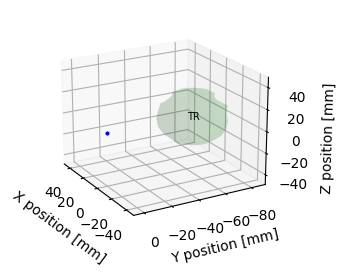}
     \label{subfig:no-cr}
  \end{minipage}}
  \hfill
  \subfloat[Small\_Mid\_CR]{
  \begin{minipage}[b]{0.30\textwidth}
     \centering
     \includegraphics[width=1.0\linewidth]{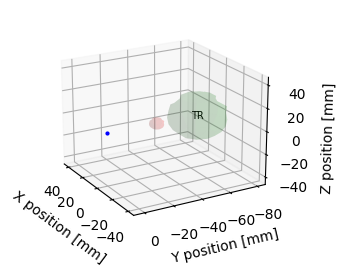}
     \label{subfig:small-mid-cr}
  \end{minipage}}
  \hfill
  \subfloat[Large\_Mid\_CR]{
  \begin{minipage}[b]{0.30\textwidth}
     \centering
     \includegraphics[width=1.0\linewidth]{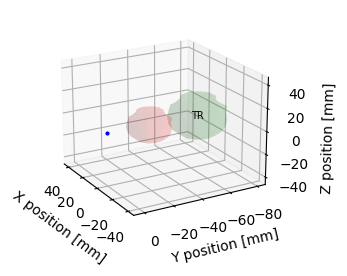}
     \label{subfig:large-mid-cr}
  \end{minipage}}
  \hfill

  \hfill
  \subfloat[Surface\_CRs]{
  \begin{minipage}[b]{0.30\textwidth}
     \centering
     \includegraphics[width=1.0\linewidth]{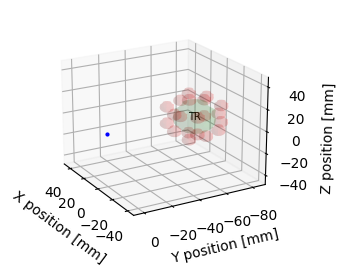}
     \label{subfig:surface-crs}
  \end{minipage}}
  \hfill
  \subfloat[Tunnel\_CRs]{
  \begin{minipage}[b]{0.30\textwidth}
     \centering
     \includegraphics[width=1.0\linewidth]{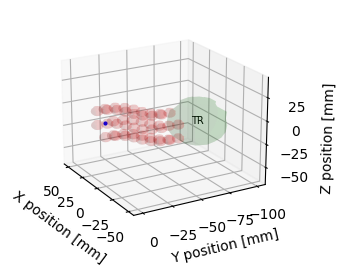}
     \label{subfig:tunnel-crs}
  \end{minipage}}
  \hfill\hspace{0.5cm}

  \caption{The environments for the online experiments (Experiment 2 and 3), each with a single TR (green) and one or more CRs (red). The dot (blue) marks the starting point.}
  \label{fig:experiment-environments}
\end{figure}

%%%%%%%%%%%%%%%%%%%%%%%%%%%%

In this section, we perform a series of experiments for the Uppaal model (to show its suitability for matching of offline measured data) as well as for the OSS framework (to show its suitability for navigation based on online measured data).
We investigate experimentally to which extent the model accuracy as well as the precision of the framework and model allow reaching TRs in simple and complex environments, and whether the performance is sufficient for a potential future use in real medical practice.
Furthermore, we check whether the experiments support the claim for global safety, i.e., that timely readjustments based on deviation and force measurements ensure that CRs are indeed never reached.

For the online experiments, we use two types of needle:
a virtual and a real needle.
The \textit{virtual needle} implements the same motion model as the Uppaal model.
Naturally, the virtual needle traces would never deviate from predicted traces, as all measurable data would be exactly known at any point.
However, to unify the handling of measured data of both the virtual and real needle, we apply the approximative fitting step of the motion circle in the virtual case as well, which then also serves as a source of artificial deviation.
The \textit{real needle} has a flexible metal shaft with a custom tip that is made from epoxy resin and beveled at an angle of approx. $45^{\circ}$.
The needle as well as the overall setup of the real experiment are shown in \cref{fig:system-setup}.
In that setting, gelatin phantoms imitate the base characteristics of real homogeneous tissue.
Compared to the virtual needle, the real needle traces can highly deviate from the predictions on a global scale, so that adjustments due to deviation are more likely in that case.

Furthermore, we use five types of environment settings, which are shown in \cref{fig:experiment-environments}.
\cref{subfig:no-cr} contains no CRs, so that reaching a TR is only affected by potential needle deviation.
\cref{subfig:small-mid-cr} and \cref{subfig:large-mid-cr} contain a single CR with small and big radius, respectively.
\cref{subfig:surface-crs} contains multiple CRs placed on the surface of the TR, and in \cref{subfig:tunnel-crs}, the CRs form a tunnel towards the TR instead.

We conduct three types of experiments:
In \textbf{Experiment 1}, we apply the Uppaal model to the data sets also used in \cite{Rogalla2020} in $\mathbb{R}^2$ (extended to $\mathbb{R}^3$ with $z=0$).
For that purpose, we measure the deviation between the measured reference traces and the set of matched simulation traces, which are determined via offline synthesis of strategies leading through targets placed on the reference trace.
In \textbf{Experiment 2}, we apply the OSS framework to the aforementioned environments and a virtual needle setup (with $50$ runs per setup), and in \textbf{Experiment 3}, we apply the subset of environments of Experiment 2 with full initial knowledge to a real needle setup (with smaller sets of $5 - 7$ runs per setup for demonstration).
Note that while the needle and tissue phantoms are real in Experiment 3, the TRs and CRs are still placed virtually at this point.
Furthermore, we abort the individual experiment runs after $120s$.
In terms of measures, recall the requirements of safety, accuracy, precision, and performance introduced in the beginning.
Safety is easily measured via the number of pierced CRs, and potential performance problems are measured by tracking the local strategy synthesis times and global experiment execution times.
Precision problems which significantly affect the verification results of the model would result in increased numbers of false positives (i.e., a critical path is determined as safe) and false negatives (i.e., existing safe paths are undetected).
False positives and negatives would then lead to a high number of readjustments and strategy resynthesis in an online setting;
false negatives would furthermore lead to a lower number of TR reachings.
Note that no exact reference model in the continuous domain exists which we could compare performance, accuracy, and precision against, as such model would not be checkable directly.
Therefore, we measure the consequences (i.e., readjustment and fewer TR reachings) as indicator for impacts on those requirements instead.

\subsection*{Evaluation and Discussion}

\begin{table}[t]
\centering
\setlength{\tabcolsep}{0.25em}
\begin{tabular}{cccccc}
\hline
\textbf{Trace} & \textbf{Deviation [mm]} & \textbf{Trace} & \textbf{Deviation [mm]} & \textbf{Trace} & \textbf{Deviation [mm]}\\
\hline
No\_Rot\_1 & $(0.53,1.35,2.04)$ & No\_Rot\_2 & $(0.92,1.72,2.35)$ & No\_Rot\_3 & $(1.05,1.58,2.00)$ \\
One\_Rot\_1 & $(1.10,1.60,2.46)$ & One\_Rot\_2 & $(1.02,1.32,2.92)$ & One\_Rot\_3 & $(0.69,1.74,4.24)$ \\
Two\_Rot\_1 & $(0.99,1.62,1.93)$ & Two\_Rot\_2 & $(0.56,1.60,3.23)$ & Two\_Rot\_3 & $(1.09,1.57,2.23)$ \\
Bad\_1 & $(1.68,2.60,4.36)$ & Bad\_2 & $(1.60,1.78,1.93)$ & & \\
\hline
\end{tabular}
\caption{The deviation (min, avg, max) between measured reference traces \cite{Rogalla2020} and matched simulation traces determined via offline strategy synthesis.}
\label{fig:trace-matching-results}
\end{table}

%%%%%%%%%%%%%%%%%%%%%%%%%%%%

\begin{table}[b]
\centering
\setlength{\tabcolsep}{0.25em}
\small
\begin{tabular}{cccccccc}
\hline
\multirow{2}{*}{\textbf{Scenario}} & \textbf{CRs} & \textbf{TR} & \textbf{CR} & \multirow{2}{*}{\textbf{Adjustments}} & \textbf{Motion} & \textbf{Synthesis} & \textbf{Overall} \\
& \textbf{Known} & \textbf{Reach} & \textbf{Hit} & & \textbf{Plans} & \textbf{Time [s]} & \textbf{Time [s]} \\
\hline
No\_CR & - & $100.00\%$ & $0.00\%$ & $(0,0.00,0)$ & $232$ & $(7.53,7.60,8.00)$ & $(18.26,24.94,34.16)$ \\
\hline
Small\_Mid\_CR & All & $100.00\%$ & $0.00\%$ & $(0,0.00,0)$ & $150$ & $(6.83,6.92,7.16)$ & $(19.05,23.42,31.51)$ \\
Small\_Mid\_CR & None & $100.00\%$ & $0.00\%$ & $(0,0.00,0)$ & $150$ & $(6.63,6.71,6.96)$ & $(18.84,23.82,33.30)$ \\
Large\_Mid\_CR & All & $0.00\%$ & $0.00\%$ & $(0,0.00,0)$ & $0$ & $(0.03,0.04,0.05)$ & $(5.17,5.18,5.20)$ \\
Large\_Mid\_CR & None & $0.00\%$ & $0.00\%$ & $(1,1.00,1)$ & $150$ & $(0.04,3.39,6.86)$ & $(19.97,20.15,20.43)$ \\
\hline
Surface\_CRs & All & $100.00\%$ & $0.00\%$ & $(0,0.06,1)$ & $27$ & $(0.10,5.67,6.96)$ & $(20.22,27.44,115.02)$ \\
Surface\_CRs & None & $40.00\%$ & $0.00\%$ & $(0,0.65,3)$ & $93$ & $(0.04,2.41,6.17)$ & $(19.30,88.03,120.00)$ \\
Tunnel\_CRs & All & $100.00\%$ & $0.00\%$ & $(0,0.16,1)$ & $166$ & $(0.09,1.85,2.31)$ & $(13.91,17.51,38.42)$ \\
Tunnel\_CRs & None & $94.00\%$ & $0.00\%$ & $(0,0.83,3)$ & $525$ & $(0.04,4.94,9.38)$ & $(20.72,34.23,110.51)$ \\
\hline
\end{tabular}
\caption{Experiment results for the virtual needle (with (min, avg, max) data for counts and times).}
\label{fig:virtual-needle-results}
\end{table}

%%%%%%%%%%%%%%%%%%%%%%%%%%%%

\begin{table}[t]
\centering
\setlength{\tabcolsep}{0.25em}
\small
\begin{tabular}{cccccccc}
\hline
\multirow{2}{*}{\textbf{Scenario}} & \textbf{CRs} & \textbf{TR} & \textbf{CR} & \multirow{2}{*}{\textbf{Adjustments}} & \textbf{Motion} & \textbf{Synthesis} & \textbf{Overall} \\
& \textbf{Known} & \textbf{Reach} & \textbf{Hit} & & \textbf{Plans} & \textbf{Time [s]} & \textbf{Time [s]} \\
\hline
No\_CR & - & $66.67\%$ & $0.00\%$ & $(1,6.00,13)$ & $469$ & $(0.05,1.14,7.87)$ & $(34.16,99.30,120.00)$ \\
\hline
Small\_Mid\_CR & All & $40.00\%$ & $0.00\%$ & $(1,1.20,2)$ & $346$ & $(0.06,2.38,9.53)$ & $(22.97,42.62,61.12)$ \\
Large\_Mid\_CR & All & $0.00\%$ & $0.00\%$ & $(0,0.00,0)$ & $0$ & $(0.04,0.06,0.09)$ & $(6.70,6.79,6.95)$ \\
\hline
Surface\_CRs & All & $57.14\%$ & $0.00\%$ & $(1,2.86,7)$ & $267$ & $(0.08,3.09,12.28)$ & $(38.18,89.25,120.00)$ \\
Tunnel\_CRs & All & $0.00\%$ & $0.00\%$ & $(0,1.33,2)$ & $317$ & $(0.07,2.45,10.49)$ & $(6.85,61.06,120.00)$ \\
\hline
\end{tabular}
\caption{Experiment results for the real needle (with (min, avg, max) data for counts and times).}
\label{fig:real-needle-results}
\end{table}

%%%%%%%%%%%%%%%%%%%%%%%%%%%%

For the experiments, we used an Ubuntu 20.04 LTS system with AMD Ryzen 7 2700X CPU and 16GB RAM, and version \textit{4.1.20-7} of \textit{Uppaal Stratego}.
The results of Experiment 1 are shown in \cref{fig:trace-matching-results}.
All normal and erroneous needle traces were covered by the model with average deviations in $1.32mm - 1.74mm$ and $1.78mm - 2.60mm$, respectively.
We notice that, while being comparable to the results of the reference work with deviations up to a few millimeters, the deviation in parts exceeds the reference values;
we assume the reason to be that the initial needle orientation was determined by another synthesis step on the model itself in \cite{Rogalla2020}, while we used the more approximative circle fitting approach.

From the results of Experiment 2 shown in \cref{fig:virtual-needle-results}, we see that the needle always reached the TR directly in the \texttt{No\_CR} scenario without readjustments.
Furthermore, we see that in all cases of known CRs, the TR is either reached directly (see readjustment counts between $0$ and $1$ and reached TR percentage of $100\%$ in \texttt{Small\_Mid\_CR}, \texttt{Surface\_CRs}, and \texttt{Tunnel\_CRs}), or the model instantly detected that the CRs block every possible way to the TR (TR reach of $0\%$ in \texttt{Large\_Mid\_CR}).
In contrast, the TR was directly reached only rarely in the case of unknown CRs (mostly in \texttt{Small\_Mid\_CR}).
Instead, most runs required a sequence of readjustments (between $1$ and $3$) before either the TR was reached (most cases of \texttt{Surface\_CRs} and \texttt{Tunnel\_CRs}) or the learned CRs blocked all ways to the TR (\texttt{Large\_Mid\_CR}), independent of whether there were actual CRs to block the way.
The average execution times of all virtual experiment setups lie between $5.18s$ and $88.03s$, and in rare cases of \texttt{Surface\_CRs}, the timeout threshold of $120s$ was reached.

Finally, the results of Experiment 3 are shown in \cref{fig:real-needle-results}.
We see clear differences to the results obtained with the virtual needle:
Using the real needle, the TR reaching rates decrease to percentages between $0\%$ (in \texttt{Tunnel\_CRs}) and $66.67\%$ (in \texttt{No\_CR}).
Furthermore, a higher number of readjustments were required (up to $13$ in one case of \texttt{No\_CR}), and thus, the average overall execution time was higher (e.g., $61.06s$ for \texttt{Tunnel\_CRs} in contrast to $17.51s$ for the virtual needle counterpart).
Yet, the CR hit percentage of $0\%$ shows that we can still prevent unsafe situations.

The results of Experiment 3 are inferior to those of Experiment 2 for a number of reasons:
The needle only approximately moves in circular paths, so that the likewise approximatively fitted motion circle may highly deviate from the actual path.
Also, the needle is deflected by formerly cut paths at times, especially after readjustment steps.
Additionally, the data acquisition process affects the results.
While the virtual needle data is exact, the real needle data depends on the type and calibration of the needle and measurement setup;
move and rotate instruction may not be executed exactly, and as pointed out in \cite{Lehmann2021}, air bubbles in gelatin and surface reflections may influence the optical measuring system.
However, it is noteworthy that even in some cases of bad local data and fitting results, the TR is still reached, and in all cases, the experiment aborts with empty strategies before reaching a potentially critical state.

%%%%%%%%%%%%%%%%%%%%%%%%%%%%%%%%%%%%%%%%%%%%%%%%%%%%%%%%%%%%%%%%%%%%%%%%%%%%%%%%
%%% Related Work %%%
%%%%%%%%%%%%%%%%%%%%%%%%%%%%%%%%%%%%%%%%%%%%%%%%%%%%%%%%%%%%%%%%%%%%%%%%%%%%%%%%
\section{Related Work} \label{sec:related-work}
Floating-point applications exist that have been successfully verified (e.g., water distribution systems \cite{Mercaldo2019} \cite{Dougherty1995} and biological kinase networks \cite{Schivo2017}) but most verification tools are restricted to discrete values.
Attempts to extend and generalize model checking by real variables were made in the past \cite{Fages2009};
however, decidability is often limited \cite{Bouajjani1995} \cite{Miller2000}, only reached via abstractions \cite{Brillout2009}, or not given at all for particular hybrid model types \cite{Henzinger1998}.
For our model, we investigate in which parts and at which points during checking the continuous system state can be described by discrete variables (to enable integer model checking) while keeping the model sufficiently accurate and precise.

Offline timed games \cite{Asarin1998} have been successfully applied to, e.g., railway, cruise, and traffic light control problems \cite{Eriksen2017}\cite{Karra2019}\cite{Basile2020} or power management \cite{Dai2017}.
These offline timed games assume complete knowledge of the behavior of a system and its environment, but few works exist that relax the requirement of complete information.
Bacci et al. investigate the partially observable oil-pump control problem and deal with imprecise knowledge of energy rates by assigning imprecisions to updates of edges to the energy-timed automata \cite{Bacci2020}.
Another approach to solve controller synthesis under partial observability is the template-based controller synthesis by Finkenbeiner and Peter, where automatic abstract refinement reduces incrementally the set of valuation of parameters until only safe states are reachable \cite{Finkbeiner2012}.
Cassez at al. \cite{Cassez2007} permit the situation where a controller strategy is generated based on (incomplete) observations but assume the strategy to be stuttering-invariant.
While most of timed games are offline approaches, in \cite{Larsen2016} Larsen at al. present a compositional online synthesis approach to control the floor heating system in a house for a short period.
Similar to our approach the floor heating system model is periodically updated with real-time sensor data of room and outside temperatures.
However, the approach assumes complete knowledge of the process, which we do not.
In the needle steering application, we cannot ignore safety-critical incomplete environmental information, namely that tissue, in reality, is inhomogeneous and has anatomic obstacles, which we are not allowed to puncture and also can only detect during the insertion.

In terms of medical technology, the use of flexible needles presents a particular research challenge: To drive flexible needles to a soft tissue target, the needle-tissue interaction needs to be known.
First approaches include nonholonomic models to estimate the needle trajectory in homogeneous tissue \cite{Webster2006}, but they do not include tissue movement due to insertion forces.
Hence, FEM modelings with linear elastic tissue have been developed \cite{Jushiddi2019}.
Yet, offline simulations of needle-tissue interaction, solely based on clinical image data, are of limited use when applied to a real clinical needle insertion as the image data does not provide details of the mechanical properties of the respective tissues.
A few adaptive online models are now available that include uncertainties from tissue inhomogeneities, needle buckling, or slip-stick-movement during needle insertion.
Those adaptive models can be applied in data-driven methods \cite{Rossa2017} or with real-time sensoring \cite{Lehmann2018}.
Still, neither method can provide any formal guarantees.

%%%%%%%%%%%%%%%%%%%%%%%%%%%%%%%%%%%%%%%%%%%%%%%%%%%%%%%%%%%%%%%%%%%%%%%%%%%%%%%%
%%% Conclusion %%%
%%%%%%%%%%%%%%%%%%%%%%%%%%%%%%%%%%%%%%%%%%%%%%%%%%%%%%%%%%%%%%%%%%%%%%%%%%%%%%%%
\section{Conclusion and Future Work} \label{sec:conclusion}
% Summary
In this paper, we designed a model and framework for online needle steering based on verifiable safety guarantees.
As the main task, we had to find proper trade-offs for the opposing requirements imposed by both the medical application and the model checker.
The experiment results have shown that the chosen set of simplifications indeed allows for safely navigating through the environment in both the virtual and real setting.

% Future Work
Still, we plan to enhance the results in three ways:
First, switching to a non-optical measuring system (e.g., ultrasound measurements) would allow mitigation of the negative impacts of gelatin phantom characteristics, and would enable the step towards application in real tissue.
Second, different needle types with varying elasticity can be used to further approach circular motions, and different needle velocities and tip sharpening might help dealing with the guiding influence of formerly cut paths.
Third, the geometric model can be replaced with a more physically accurate one. Depending on the concrete use case, one may implement additional motion restrictions to either simultaneously relax some of the current restrictions (e.g., the limited number of allowed rotation per synthesis phase), or implement a more dynamic environment model of heterogeneous tissue with moving CRs and TRs.
Finally, we note that the underlying OSS framework can be applied to other navigation tasks beyond needle steering.

%%%%%%%%%%%%%%%%%%%%%%%%%%%%%%%%%%%%%%%%%%%%%%%%%%%%%%%%%%%%%%%%%%%%%%%%%%%%%%%%
%%% BIBLIOGRAPHY %%%
%%%%%%%%%%%%%%%%%%%%%%%%%%%%%%%%%%%%%%%%%%%%%%%%%%%%%%%%%%%%%%%%%%%%%%%%%%%%%%%%
% \newpage
\bibliographystyle{eptcs}
\bibliography{bib/references}
% \nocite{*}

%%%%%%%%%%%%%%%%%%%%%%%%%%%%%%%%%%%%%%%%%%%%%%%%%%%%%%%%%%%%%%%%%%%%%%%%%%%%%%%%
%%% APPENDIX %%%
%%%%%%%%%%%%%%%%%%%%%%%%%%%%%%%%%%%%%%%%%%%%%%%%%%%%%%%%%%%%%%%%%%%%%%%%%%%%%%%%
\appendix
\section{C Code Declaration of the Uppaal Model} \label{app:c-code-declaration}
The following code section presents the complete C code declaration of the Uppaal needle steering model:
\begin{lstlisting}[
  basicstyle=\ttfamily\tiny,
  language=C,
  frame=single
  ]
typedef int[-(1<<31),(1<<31)-1] int32_t;   // int32_t: A 32-bit integer data type.

const int32_t INT32_MIN = -2147483648;     // INT32_MIN: The minimum value of the 32-bit integer range.
const int32_t INT32_MAX = 2147483647;      // INT32_MAX: The maximum value of the 32-bit integer range.

const int _SI_ = 1000;                     // _SI_: The scale factor between integer and double domain (as integer).
const double PI = 3.14159265358979312;     // PI: The (rounded) value of pi.

int dt = 2;                                // dt: The simulation time delta.

/*************************************************
************ General Helper Functions ************
*************************************************/
/**
 * @brief Determines the maximum of two given values.
 * @param v1 The first value.
 * @param v2 The second value.
*/
int32_t max(int32_t v1, int32_t v2) {
    return (v1 > v2 ? v1 : v2);
}

/**
 * @brief Determines the minimum of two given values.
 * @param v1 The first value.
 * @param v2 The second value.
*/
int32_t min(int32_t v1, int32_t v2) {
    return (v1 < v2 ? v1 : v2);
}

/*****************************************************
******************** System Model ********************
*****************************************************/

/****************
* Actions (ACT) *
****************/
const int32_t MIN_PROG_BETWEEN_ROTS = 2*_SI_;  // MIN_PROG_BETWEEN_ROTS: The required needle progress between rotations.
const int MAX_ROT_COUNT = 2;                   // MAX_ROT_COUNT: The maximum allowed number of rotations.

const int PUSH_ACTION = 0;                     // PUSH_ACTION: The ID of a push action.
const int ROLL_ROTATE_ACTION = 1;              // ROLL_ROTATE_ACTION: The ID of a roll rotate action.

const int MAX_ACT_COUNT = 10;                  // MAX_ACT_COUNT: The maximum allowed number of actions.
const int ACT_PROG = 0;                        // ACT_PROG: The index of the needle progress inside an action tuple.
const int ACT_TYPE = 1;                        // ACT_TYPE: The index of the action type inside an action tuple.
const int ACT_VAL = 2;                         // ACT_VAL: The index of the action value inside an action tuple.
int32_t actions[MAX_ACT_COUNT][3];             // Tuples (progress, action_type, action_value).
int32_t curr_ACT_index = 0;                    // curr_ACT_index: The current action index.

int ROT_counter = 0;                           // ROT_counter: A counter for the number of rotations.
int prev_ROT_index = -1;                       // prev_ROT_index: The index of the most previous rotation action.

const int ANGLE_COUNT = 3;                     // ANGLE_COUNT: The count of different angles for rotation actions.
int allowed_rot_angles[ANGLE_COUNT] = {        // allowed_rot_angles: The set of allowed rotation angle.
  90, 180, 270
};
int sel_rot_angle = 0;                         // sel_rot_angle: The currently selected rotation angle.

/*************************
* Detection Regions (DR) *
*************************/
const int32_t DR_COUNT = 1;                  // DR_COUNT: The number of detection regions in the system.

const bool HAS_DETECTION_REGIONS = true;     // HAS_DETECTION_REGIONS: Indicator for whether the system has any DRs.
const bool TREAT_DR_AS_CR = true;            // TREAT_DR_AS_CR: Indicator for whether DRs should be handle similar to
                                             //   CRs, i.e., paths that reach DRs are discarded as well.
int32_t DR_center[DR_COUNT][3] = {           // DR_center: The center points of the spherical DRs.
    {0, -25*_SI_, 0}
};
int32_t DR_radius[DR_COUNT] = { 5*_SI_ };    // DR_radius: The radii of the spherical DRs.

/************************
* Critical Regions (CR) *
************************/
const int32_t CR_COUNT = 1;                  // CR_COUNT: The number of critical regions in the system.

const bool HAS_CRITICAL_REGIONS = true;      // HAS_CRITICAL_REGIONS: Indicator for whether the system has any CRs.
int32_t CR_center[CR_COUNT][3] = {           // CR_center: The center points of the spherical CRs.
    {0, -25*_SI_, 0}
};
int32_t CR_radius[CR_COUNT] = { 2*_SI_ };    // CR_radius: The radii of the spherical CRs.

/**********************
* Target Regions (TR) *
**********************/
const int32_t TR_COUNT = 1;                  // TR_COUNT: The number of target regions in the system.

int32_t TR_center[TR_COUNT][3] = {           // TR_center: The center points of the spherical TRs.
    {0, -50*_SI_, 0}
};
int32_t TR_radius[TR_COUNT] = { 10*_SI_ };   // TR_radius: The radii of the spherical TRs.

int32_t curr_TR_index = 0;                   // curr_TR_index: The index of the currently considered TR.

/*****************************************************
******************** Needle Model ********************
*****************************************************/

/*****************************
* Needle Parameters and Data *
*****************************/
int32_t needle_progress = 0;                     // needle_progress: The current progress of the needle [mm].
int32_t tip_pos[3] = {0*_SI_, 0*_SI_, 0*_SI_};   // tip_pos: The current position of the needle tip [mm].
int32_t needle_vel = 1 * _SI_;                   // needle_vel: The velocity of the needle [mm/s].
int32_t tip_roll = 0 * _SI_;                     // tip_roll: The current roll angle of the needle tip.

/** Motion Circle **/
int32_t motion_circle_center[3] = {              // motion_circle_center: The center of the current motion circle.
  0*_SI_, 0*_SI_, 0*_SI_
};
int32_t motion_circle_r = 30 * _SI_;             // motion_circle_r: The radius of the current motion circle.
int32_t motion_circle_yaw = 0 * _SI_;            // motion_circle_yaw: The current tip yaw around the motion circle.

/** Vectors **/
const int GLOBAL_X_VEC = 0;   // GLOBAL_X_VEC: The ID of the x-axis of the global coordinate system.
const int GLOBAL_Y_VEC = 1;   // GLOBAL_Y_VEC: The ID of the y-axis of the global coordinate system.
const int GLOBAL_Z_VEC = 2;   // GLOBAL_Z_VEC: The ID of the z-axis of the global coordinate system.
const int TIP_X_VEC = 3;      // TIP_X_VEC: The ID of the x-axis of the local coordinate system of the needle tip.
                              //   -> pointing in needle motion direction.
const int TIP_Y_VEC = 4;      // TIP_X_VEC: The ID of the y-axis of the local coordinate system of the needle tip.
                              //   -> pointing towards rotation circle center.
const int TIP_Z_VEC = 5;      // TIP_X_VEC: The ID of the z-axis of the local coordinate system of the needle tip.
                              //   -> normal of rotation circle.
const int TEMP_VEC_1 = 6;     // TEMP_VEC_1: A vector for intermediate calculation results.
const int TEMP_VEC_2 = 7;     // TEMP_VEC_2: A vector for intermediate calculation results.

const int VECTOR_COUNT = 8;
int32_t vectors[VECTOR_COUNT][3] = {
    {1*_SI_, 0, 0},   // GLOBAL_X_VEC
    {0, 1*_SI_, 0},   // GLOBAL_Y_VEC
    {0, 0, 1*_SI_},   // GLOBAL_Z_VEC
    {0, 0, 0},        // TIP_X_VEC
    {0, 0, 0},        // TIP_Y_VEC
    {0, 0, 0},        // TIP_Z_VEC
    {0, 0, 0},        // TEMP_VEC_1
    {0, 0, 0}         // TEMP_VEC_2
};

/** Matrices **/
const int ID_MAT = 0;               // ID_MAT: The identity matrix.
const int TIP_BASE_ROT_MAT = 1;     // TIP_BASE_ROT_MAT: Represents needle rotation after last "motion -> rotation".
const int TIP_CIRCLE_ROT_MAT = 2;   // TIP_CIRCLE_ROT_MAT: Reprents the rotation caused by the circular motion.
const int TIP_ACTION_ROT_MAT = 3;   // TIP_ACTION_ROT_MAT: Reprents the rotation caused by a needle roll action.
const int TIP_ROT_MAT = 4;          // TIP_ROT_MAT: Represents the current rotation of the needle tip.
const int TEMP_MAT_1 = 5;           // TEMP_MAT_1: A matrix for intermediate calculation results.

const int MATRIX_COUNT = 6;
int32_t matrices[MATRIX_COUNT][3][3] = {
    {{1*_SI_, 0, 0}, {0, 1*_SI_, 0}, {0, 0, 1*_SI_}},   // ID_MAT
    {{0, 1*_SI_, 0}, {-1*_SI_, 0, 0}, {0, 0, 1*_SI_}},  // TIP_BASE_ROT_MAT
    {{1*_SI_, 0, 0}, {0, 1*_SI_, 0}, {0, 0, 1*_SI_}},   // TIP_CIRCLE_ROT_MAT
    {{1*_SI_, 0, 0}, {0, 1*_SI_, 0}, {0, 0, 1*_SI_}},   // TIP_ACTION_ROT_MAT
    {{1*_SI_, 0, 0}, {0, 1*_SI_, 0}, {0, 0, 1*_SI_}},   // TIP_ROT_MAT
    {{0, 0, 0}, {0, 0, 0}, {0, 0, 0}}                   // TEMP_MAT_1
};


/****************************
* Vector / matrix functions *
****************************/
/**
 * @brief Calculates the dot product of two matrices.
 * @param in_mat_1 The index of the first input matrix.
 * @param in_mat_2 The index of the second input matrix.
 * @param out_mat The index of the resulting matrix.
*/
void matrix_matrix_dot(int in_mat_1, int in_mat_2, int out_mat) {
    int i,j,k;
    int32_t entry_sum;
    double _S_ = _SI_*1.0;  // _S_: The scale factor (as double value).

    for (i=0; i<3; i++) {
        for (j=0; j<3; j++) {
            entry_sum = 0;
            for (k=0; k<3; k++) {
                entry_sum += fint(((matrices[in_mat_1][i][k]/_S_) * (matrices[in_mat_2][k][j]/_S_))*_S_);
            }
            matrices[out_mat][i][j] = entry_sum;
        }
    }
}

/**
 * @brief Calculates the dot product of a matrix and a vector.
 * @param in_mat The index of the input matrix.
 * @param in_vec The index of the input vector.
 * @param out_vec The index of the resulting vector.
*/
void matrix_vector_dot(int in_mat, int in_vec, int out_vec) {
    int i,k;
    int32_t entry_sum;
    double _S_ = _SI_*1.0;  // _S_: The scale factor (as double value).

    for (i=0; i<3; i++) {
        entry_sum = 0;
        for (k=0; k<3; k++) {
            entry_sum += fint(((matrices[in_mat][i][k]/_S_) * (vectors[in_vec][k]/_S_))*_S_);
        }
        vectors[out_vec][i] = entry_sum;
    }
}

/**
 * @brief Copies a vector to another vector.
 * @param in_vec The index of the source vector.
 * @param out_vec The index of the target vector.
*/
void copy_vector(int in_vec, int out_vec) {
    int i;
    for (i=0; i<3; i++) {
        vectors[out_vec][i] = vectors[in_vec][i];
    }
}

/**
 * @brief Copies a matrix to another matrix.
 * @param in_mat The index of the source matrix.
 * @param out_mat The index of the target matrix.
*/
void copy_matrix(int in_mat, int out_mat) {
    int i,j;
    for (i=0; i<3; i++) {
        for (j=0; j<3; j++) {
            matrices[out_mat][i][j] = matrices[in_mat][i][j];
        }
    }
}

/**
 * @brief Get the magnitude of a given vector.
 * @param in_vec The index of the vector.
*/
int32_t get_magnitude(int in_vec) {
    double _S_ = _SI_*1.0;  // _S_: The scale factor (as double value).

    int32_t res = fint((sqrt(pow(vectors[in_vec][0]/_S_, 2) +
                             pow(vectors[in_vec][1]/_S_, 2) +
                             pow(vectors[in_vec][2]/_S_, 2)))*_S_);
    return res;
}

/**
 * @brief Normalizes a given vector.
 * @param in_vec The index of the vector.
*/
void normalize_vector(int in_vec) {
    int i;
    double _S_ = _SI_*1.0;  // _S_: The scale factor (as double value).

    int32_t magnitude = get_magnitude(in_vec);
    if (magnitude == 0) {
        return;
    }
    for (i=0; i<3; i++) {
        vectors[in_vec][i] = fint(((vectors[in_vec][i]/_S_) / (magnitude/_S_)) * _S_);
    }
}

/**
 * @brief Generates a rotation matrix based on an axis vector and an angle.
 * @param axis The index of the axis vector.
 * @param theta The angle.
 * @param out_mat The index of the resulting matrix.
*/
void generate_rotation_matrix_from_axis_and_angle(int axis, int32_t theta, int out_mat) {
    double _S_ = _SI_*1.0;  // _S_: The scale factor (as double value).

    matrices[out_mat][0][0] = fint((cos(theta/_S_) + pow(vectors[axis][0]/_S_,2) * (1 - cos(theta/_S_)))*_S_);
    matrices[out_mat][0][1] = fint(((vectors[axis][0]/_S_) * (vectors[axis][1]/_S_) *
                                (1 - cos(theta/_S_)) - (vectors[axis][2]/_S_) * sin(theta/_S_))*_S_);
    matrices[out_mat][0][2] = fint(((vectors[axis][0]/_S_) * (vectors[axis][2]/_S_) *
                                (1 - cos(theta/_S_)) + (vectors[axis][1]/_S_) * sin(theta/_S_))*_S_);
    matrices[out_mat][1][0] = fint(((vectors[axis][1]/_S_) * (vectors[axis][0]/_S_) *
                                (1 - cos(theta/_S_)) + (vectors[axis][2]/_S_) * sin(theta/_S_))*_S_);
    matrices[out_mat][1][1] = fint((cos(theta/_S_) + pow(vectors[axis][1]/_S_,2) * (1 - cos(theta/_S_)))*_S_);
    matrices[out_mat][1][2] = fint(((vectors[axis][1]/_S_) * (vectors[axis][2]/_S_) *
                                (1 - cos(theta/_S_)) - (vectors[axis][0]/_S_) * sin(theta/_S_))*_S_);
    matrices[out_mat][2][0] = fint(((vectors[axis][2]/_S_) * (vectors[axis][0]/_S_) *
                                (1 - cos(theta/_S_)) - (vectors[axis][1]/_S_) * sin(theta/_S_))*_S_);
    matrices[out_mat][2][1] = fint(((vectors[axis][2]/_S_) * (vectors[axis][1]/_S_) *
                                (1 - cos(theta/_S_)) + (vectors[axis][0]/_S_) * sin(theta/_S_))*_S_);
    matrices[out_mat][2][2] = fint((cos(theta/_S_) + pow(vectors[axis][2]/_S_,2) * (1 - cos(theta/_S_)))*_S_);
}

/**********************
* Simulator functions *
**********************/
/**
 * @brief Updates the tip rotation matrix (composed from base, circle, and action rotation).
*/
void update_tip_rotation_matrix() {
    matrix_matrix_dot(TIP_CIRCLE_ROT_MAT, TIP_BASE_ROT_MAT, TEMP_MAT_1);
    matrix_matrix_dot(TIP_ACTION_ROT_MAT, TEMP_MAT_1, TIP_ROT_MAT);
}

/**
 * @brief Updates the action rotation matrix (e.g., when an action was performed).
*/
void update_action_rotation_matrix() {
    generate_rotation_matrix_from_axis_and_angle(TIP_X_VEC, tip_roll, TIP_ACTION_ROT_MAT);
}

/**
 * @brief Updates the circle rotation matrix (i.e., when the needle was further moved along the rotation circle).
*/
void update_circle_rotation_matrix() {
    matrix_vector_dot(TIP_BASE_ROT_MAT, GLOBAL_Y_VEC, TEMP_VEC_1);
    normalize_vector(TEMP_VEC_1);
    generate_rotation_matrix_from_axis_and_angle(TEMP_VEC_1, motion_circle_yaw, TIP_CIRCLE_ROT_MAT);
}

/**
 * @brief Reset the remaining matrices (circle and action rotation) and the rotation circle yaw
 *        after base rotation update.
*/
void reset_data_after_base_rotation_update() {
    copy_matrix(ID_MAT, TIP_CIRCLE_ROT_MAT);
    copy_matrix(ID_MAT, TIP_ACTION_ROT_MAT);
    motion_circle_yaw = 0;
}

/**
 * @brief Updates the tip base rotation matrix (composed from base, circle, and action rotation).
          The current, concrete tip rotation matrix becomes the new base rotation matrix then.
          All other matrices (circle and action rotation) and the rotation circle yaw are reset afterwards.
*/
void update_base_rotation_matrix() {
    update_tip_rotation_matrix();
    copy_matrix(TIP_ROT_MAT, TIP_BASE_ROT_MAT);

    reset_data_after_base_rotation_update();
}

/**
 * @brief Updates the local coordinate system of the needle.
 *        The x-axis points in needle direction, the y-axis towards the current rotation circle center.
*/
void update_needle_coordinate_system() {
    matrix_vector_dot(TIP_ROT_MAT, GLOBAL_X_VEC, TIP_X_VEC);
    normalize_vector(TIP_X_VEC);
    matrix_vector_dot(TIP_ROT_MAT, GLOBAL_Y_VEC, TIP_Y_VEC);
    normalize_vector(TIP_Y_VEC);
    matrix_vector_dot(TIP_ROT_MAT, GLOBAL_Z_VEC, TIP_Z_VEC);
    normalize_vector(TIP_Z_VEC);
}

/**
 * @brief Updates the rotation circle center (e.g., when the needle roll angle changes).
*/
void update_circle_center() {
    double _S_ = _SI_*1.0;  // _S_: The scale factor (as double value).

    update_tip_rotation_matrix();
    update_needle_coordinate_system();
    motion_circle_center[0] = fint((tip_pos[0]/_S_ - ((motion_circle_r/_S_) * (vectors[TIP_Z_VEC][0]/_S_))) * _S_);
    motion_circle_center[1] = fint((tip_pos[1]/_S_ - ((motion_circle_r/_S_) * (vectors[TIP_Z_VEC][1]/_S_))) * _S_);
    motion_circle_center[2] = fint((tip_pos[2]/_S_ - ((motion_circle_r/_S_) * (vectors[TIP_Z_VEC][2]/_S_))) * _S_);
}

/**
 * @brief Updates the position of the needle tip.
*/
void update_tip_pos() {
    double _S_ = _SI_*1.0;  // _S_: The scale factor (as double value).

    update_circle_rotation_matrix();

    // Calculate rotated vector from circle center to new tip position
    matrix_vector_dot(TIP_BASE_ROT_MAT, GLOBAL_Z_VEC, TEMP_VEC_1);
    matrix_vector_dot(TIP_CIRCLE_ROT_MAT, TEMP_VEC_1, TEMP_VEC_2);

    // Calculate new tip position
    tip_pos[0] = fint((motion_circle_center[0]/_S_ + (vectors[TEMP_VEC_2][0]/_S_) * (motion_circle_r/_S_)) * _S_);
    tip_pos[1] = fint((motion_circle_center[1]/_S_ + (vectors[TEMP_VEC_2][1]/_S_) * (motion_circle_r/_S_)) * _S_);
    tip_pos[2] = fint((motion_circle_center[2]/_S_ + (vectors[TEMP_VEC_2][2]/_S_) * (motion_circle_r/_S_)) * _S_);

    update_tip_rotation_matrix();
    update_needle_coordinate_system();
}

/**
 * @brief Rotates the needle by a given roll angle specified in radians
          (e.g., rotate_rad(fint(0.5*PI*_SI_)) for a 90 degree rotation).
*/
void rotate_rad(int32_t angle) {
    tip_roll = angle;
    update_action_rotation_matrix();
    update_circle_center();

    // Update tip base rotation matrix by previous base rotation, circle rotation, and action rotation
    update_base_rotation_matrix();

    ROT_counter++;
}

/**
 * @brief Rotates the needle by a given roll angle specified in degrees
          (e.g., rotate_deg(90) for a 90 degree rotation).
*/
void rotate_deg(int32_t angle) {
    rotate_rad(fint(((angle/180.0) * PI * _SI_)));
}

/**
 * @brief Performs a single simulation step.
*/
void step() {
    int32_t needle_pos_delta, tip_yaw_delta;
    double _S_ = _SI_*1.0;  // _S_: The scale factor (as double value).

    needle_pos_delta = fint(((needle_vel/_S_) * dt) * _S_);
    tip_yaw_delta = fint(((needle_pos_delta/_S_) / (motion_circle_r/_S_)) * _S_);

    needle_progress += needle_pos_delta;
    motion_circle_yaw += tip_yaw_delta;
    update_tip_pos();
}

/**
 * @brief Initializes the needle.
*/
void initialize() {
    update_circle_center();
}

/*************************************************
******************** TA Model ********************
*************************************************/

/** Clocks **/
clock t = 0.0;  // t: The global system clock.

/** Communication Channels **/
/* Workflow */
broadcast chan do_tissue_action;         // do_tissue_action: The channel for notifying the Tissue that needle action
                                         //   step is finished to start Tissue reaction.
broadcast chan finished_tissue_action;   // finished_tissue_action: The channel for notifying the User that Tissue
                                         //   reaction has finished.
broadcast chan do_check;                 // do_check: The channel for notifying the State Checker that a new state is
                                         //   available for validity / reaching checks.
broadcast chan finished_check;           // finished_check: The channel for notifying the User that the State Checker
                                         //   finished the validity / reaching checks.

/* User / Needle */
broadcast chan do_push;     // do_push: The channel for notifying the Needle that a push motion should be executed.
broadcast chan do_rotate;   // do_rotate: The channel for notifying the Needle that a rotate motion should be executed.

/* Checker */
broadcast chan terminate;   // terminate: The channel for notifying the User that the current path should be discarded.

/*******************
* Logger functions *
*******************/
/**
 * @brief Logs the current push action.
*/
void log_push_action() {
    actions[curr_ACT_index][ACT_PROG] = curr_ACT_index==0 ? 0 : actions[curr_ACT_index-1][ACT_PROG];
    actions[curr_ACT_index][ACT_TYPE] = PUSH_ACTION;
    actions[curr_ACT_index][ACT_VAL] = needle_progress - actions[curr_ACT_index][ACT_PROG];
    curr_ACT_index++;
}

/**
 * @brief Logs the current rotation action.
*/
void log_rotation_action() {
    actions[curr_ACT_index][ACT_PROG] = needle_progress;
    actions[curr_ACT_index][ACT_TYPE] = ROLL_ROTATE_ACTION;
    actions[curr_ACT_index][ACT_VAL] = sel_rot_angle;
    prev_ROT_index = curr_ACT_index;
    curr_ACT_index++;

}

/******************************
* Guard / Invariant functions *
******************************/
/**
 * @brief Checks if the needle tip entered any critical region.
*/
bool CR_reached = false;

bool check_CR_reached() {
    int32_t tip_center_dist, tip_region_dist;
    int i;
    double _S_ = _SI_*1.0;
    CR_reached = false;

    if (!HAS_CRITICAL_REGIONS && !HAS_DETECTION_REGIONS) {
        CR_reached = false;
        return CR_reached;
    }

    for (i=0; i<CR_COUNT; i++) {
        tip_center_dist = fint(sqrt(
            pow(CR_center[i][0]/_S_ - tip_pos[0]/_S_,2) +
            pow(CR_center[i][1]/_S_ - tip_pos[1]/_S_,2) +
            pow(CR_center[i][2]/_S_ - tip_pos[2]/_S_,2))*_S_);
        tip_region_dist = tip_center_dist - CR_radius[i];
        if (tip_region_dist <= 0) {
            CR_reached = true;
            return CR_reached;
        }
    }

    if (TREAT_DR_AS_CR) {
        if (!HAS_DETECTION_REGIONS) {
            CR_reached = false;
            return CR_reached;
        }

        for (i=0; i<DR_COUNT; i++) {
            tip_center_dist = fint(sqrt(
                pow(DR_center[i][0]/_S_ - tip_pos[0]/_S_,2) +
                pow(DR_center[i][1]/_S_ - tip_pos[1]/_S_,2) +
                pow(DR_center[i][2]/_S_ - tip_pos[2]/_S_,2))*_S_);
            tip_region_dist = tip_center_dist - DR_radius[i];
            if (tip_region_dist <= 0) {
                CR_reached = true;
                return CR_reached;
            }
        }
    }

    return CR_reached;
}

/**
 * @brief Checks if the needle tip lies within a defined target region.
*/
bool TR_reached = false;
bool check_specific_TR_reached(int TR_index) {
    double tip_center_dist, tip_region_dist;
    TR_reached = false;
    tip_center_dist = sqrt(pow(TR_center[TR_index][0] - tip_pos[0],2) +
                           pow(TR_center[TR_index][1] - tip_pos[1],2) +
                           pow(TR_center[TR_index][2] - tip_pos[2],2));
    tip_region_dist = tip_center_dist - TR_radius[TR_index];
    if (tip_region_dist <= 0) {
        TR_reached = true;
    }
    return TR_reached;
}
bool check_TR_reached() {
    return check_specific_TR_reached(curr_TR_index);
}

/**
 * @brief Checks if the needle tip can still reach the target region.
*/
bool TR_reachable = false;
bool check_specific_TR_reachable(int TR_index) {
    TR_reachable = (tip_pos[1] >= (TR_center[TR_index][1] - TR_radius[TR_index]));
    return TR_reachable;
}
bool check_TR_reachable() {
    return check_specific_TR_reachable(curr_TR_index);
}

/**
 * @brief Checks if the current needle path is still valid
          (i.e., if the needle moves in a valid direction).
*/
bool path_valid = false;
bool check_path_valid() {
    bool is_negative_y_motion = vectors[TIP_X_VEC][1] <= 0;
    path_valid = is_negative_y_motion;
    return path_valid;
}
\end{lstlisting}

\end{document}